\documentclass[12pt]{article}
\usepackage{graphicx,epsfig,color}
\newcommand{\bc}{\begin{center}}
\newcommand{\ec}{\end{center}}
\newcommand{\be}{\begin{equation}}
\newcommand{\ee}{\end{equation}}
\newcommand{\bea}{\begin{eqnarray}}
\newcommand{\eea}{\end{eqnarray}}
\newcommand{\ba}{\begin{array}}
\newcommand{\ea}{\end{array}}
\newcommand{\lb}{\label}
\newcommand{\rf}{\ref}
\newcommand{\bfg}{\begin{figure}[htbp]}
\newcommand{\efg}{\end{figure}}
\newcommand{\pr}{Phys. Rev. }
\newcommand{\prd}{Phys. Rev. D }
\newcommand{\np}{Nucl. Phys. }
\newcommand{\npb}{Nucl. Phys. B }
\newcommand{\prl}{Phys. Rev. Lett. }
\newcommand{\prp}{Phys. Rep. }
\newcommand{\ap}{Ann. Phys. (N.Y.) }
\newcommand{\pl}{Phys. Lett. }
\newcommand{\plb}{Phys. Lett. B }

\newcommand{\jp}{J. Phys. }
\newcommand{\rmp}{Rev. Mod. Phys. }

\makeatletter
\AtBeginDocument{\@ifpackageloaded{natbib}{\ifNAT@numbers\if@filesw\immediate\write\@auxout{\string\global\string\NAT@numberstrue}\fi\fi}{}}
\makeatother
\begin{document}

\vspace*{1. cm}
\bc
{\large \textbf{Gauge-invariant approach to quark dynamics}}
\\
\vspace{1 cm}
H. Sazdjian\\
Institut de Physique Nucl\'eaire, CNRS-IN2P3,\\
Universit\'e Paris-Sud, Universit\'e Paris-Saclay, 91406 Orsay Cedex, France\\
E-mail: sazdjian@ipno.in2p3.fr
\ec
\par
\renewcommand{\thefootnote}{\fnsymbol{footnote}}
\vspace{0.75 cm}

\bc
{\large Abstract}
\ec
\par
The main aspects of a gauge-invariant approach to the description of
quark dynamics in the nonperturbative regime of QCD are first reviewed. 
In particular, the role of the parallel transport operation in 
constructing gauge-invariant Green's functions is presented, and the 
relevance of Wilson loops for the representation of the interaction 
is emphasized.
Recent developments, based on the use of polygonal lines for the parallel 
transport operation, are then presented.
An integ\-ro\--differen\-tial equation is obtained for the quark Green's 
function defined with a phase factor along a single, straight line segment.  
It is solved exactly and analytically in the case of two-dimensional QCD 
in the large-$N_c$ limit. The solution displays the dynamical mass 
generation phenomenon for quarks, with an infinite number of branch-cut 
singularities that are stronger than simple poles.
\par  
\vspace{0.5 cm}
\par

\newpage

\section{Introduction} \lb{s1}
\setcounter{equation}{0}

Quantum chromodynamics (QCD), the theory of the strong interaction
in which the fundamental fields are the quarks (the matter fields)
and the gluons (the gauge fields), has the primary property of being
asymptotically free \cite{Gross:1973id,Politzer:1973fx}: The 
interaction weakens at short distances between particles or sources,
a feature that allows the use of perturbation theory in that domain. 
This theoretical prediction has been widely verified by many experimental 
processes and data and represents one of the major justifications for
the advent of QCD.
\par
The counterpart of asymptotic freedom is that the effective coupling
constant increases at large distances, where perturbation theory 
ceases to be valid. The failure of perturbation theory in the infrared 
domain actually  has a deep origin, confirmed by nature: Quarks and 
gluons, the fundamental constituents of QCD, are not observed in nature 
as asymptotic free particles; at low energies or at large-distance 
separations, they are confined into color singlet bound states, which 
are the hadrons (mesons and baryons). Their presence there can only  be
detected indirectly through spectroscopic analyses. This phenomenon 
is a new feature that had not been addressed in earlier known field 
theories. 
\par
The precise description of the confinement mechanism requires a 
nonperturbative approach and still remains a challenge on analytic 
grounds. 
\par
The only systematic tool that succeeds in enabling the treatment of the 
large-distance regime of QCD is lattice theory, first introduced
by Wilson \cite{Wilson:1974sk,Kogut:1982ds,Kaku:1993ym}. Lattice theory 
is based on the discretization of spacetime and a numerical calculation 
of the path integral. Its limitations come from the facts that the theory
is considered in Euclidean space, rather than Minkowski space, and the
continuous Poincar\'e invariance is explicitly broken by the 
discretization mechanism and replaced by discrete symmetries. 
Furthermore,  limitations of computational power  also place 
restrictions on the precision of the calculation. It is worthwhile 
noticing, to the credit of lattice QCD, that it predicts analytically, 
in the strong-coupling approximation, the confinement of quarks.
\par
Another systematic tool is provided by the Dyson--Schwinger equations
\cite{Dyson:1949ha,Schwinger:1951ex,Alkofer:2000wg,Fischer:2006ub}.       
These are integral equations for Green's functions of the theory,
which, however, are infinite in number and coupled to each other. 
For a practical resolution, one considers mainly the equations relative
to the simplest Green's functions with a truncation of the infinite
series of coupled sectors. Numerical solutions compatible with 
confinement have been found. Also, solutions verifying chiral symmetry
breaking have been obtained \cite{Maris:1997hd}. 
This line of approach is still under investigation. 
Inherent difficulties might be related with the gauge-invariance problem 
in nonperturbative approaches, where truncation 
of exact equations should be done in a consistent way in order not
to break gauge invariance of physical quantities to the order of
approximation that is considered. In QCD, because of the masslessness
of the gluon field, artificial infrared divergences or singularities 
easily appear in intermediate quantities. Ensuring infrared finiteness
of physical quantities is another difficult task in nonperturbative approaches.
\par   
Because of the infrared instabilities of noninvariant quantities under
gauge transformations, a line of approach based on the use from the 
start of gauge-invariant quantities has been under development for a long 
time \cite{Mandelstam:1962mi,Mandelstam:1962us,BialynickiBirula:1963,
Mandelstam:1968hz,Nambu:1978bd}. However, because gauge-invariant 
Green's functions are extended composite objects, equations derived 
for them have more complicated structure and this fact has inhibited  
rapid and straightforward progress. Nevertheless, it turns out that 
the confinement problem has a sounder formulation in this approach 
than in more conventional approaches and provides simple criteria for 
its verification \cite{Wilson:1974sk,Brown:1979ya}.
\par
The remaining part of this article is devoted to a presentation of 
characteristic features of the corresponding formalism and of new
developments obtained in recent years.
\par 

\section{Gauge transformations} \lb{s2}

The gauge group of QCD is SU$(N_c)$, acting in the internal space of 
color degrees of freedom; the physical value of $N_c$ is 3, but leaving 
$N_c$ as a free parameter allows one to study in more generality the 
properties of the theory.
\par
The quark fields belong to the defining ($N_c$-dimensional) fundamental 
representation of the gauge group. They are denoted $\psi_{\alpha}^a(x)$, 
where $a$, the color index, runs from 1 to $N_c$; $\alpha$ is the spinor 
index; and $x$ is the spacetime coordinate, which we shall consider in 
Minkowski space. Antiquark fields are denoted 
$\overline{\psi}_{b,\beta}(x)$; they belong to the conjugate 
($N_c$-dimensional) fundamental representation.
Actually, there are six different types of quark, with different masses 
and appropriate electric charges, classified according to a global 
index, called flavor; however, for simplicity, we shall consider in
the following only one type of quark, without specifying its flavor.
(In QCD, the interaction does not couple different flavor states.)
Under a local gauge transformation, the quark and antiquark fields 
transform as
\be \lb{e1}
\psi_{\alpha}^a(x)\longrightarrow \psi_{\alpha}^{'a}(x)=
\Omega_{\ c}^a(x)\psi_{\alpha}^c(x),
\ \ \ \ \ \overline{\psi}_{b,\beta}(x)\longrightarrow 
\overline{\psi}_{b,\beta}^{\ '}(x)=
\overline{\psi}_{c,\beta}(x)\Omega_{\ \ b}^{\dagger c}(x),
\ee
where $\Omega$ is an element of the group SU$(N_c)$, whose parameters
are in general $x$ dependent (unless specified otherwise, repeated 
indices are assumed to be summed over).  
\par
Because of the $x$ dependence of the parameters of the group 
transformations, the derivatives of the quark and antiquark fields
do not transform as members of the fundamental representations, as
in Eqs. (\rf{e1}). Fields or functions of fields that transform as 
being members of irreducible representations will be said to be transforming 
covariantly, although here the term covariance is not related to the 
position of the color index (upper or lower).
\par
To define a covariant derivative operator one introduces
a connection, represented by the gluon gauge field, that belongs,
for $x$-independent gauge transformations, to the adjoint representation
of the group; since the latter can be obtained from the composition of 
the fundamental representation with its conjugate, the gluon field can
be represented as depending on two color indices, related to the two
previous representations: $A_{\ b,\mu}^a$ \cite{'tHooft:1973jz}.
One also has the properties $A_{\ \ b,\mu}^{\dagger a}=A_{\ a,\mu}^b$ 
and $A_{\ a,\mu}^a=0$.
The more conventional notation represents the gluon field with a
single index $B$ running from 1 to $(N_c^2-1)$. The relationship between 
the two notations can be obtained in the following way. If $T^B$ are the
generators of the group transformations in the fundamental representation,   
then we have $(T^BA_{\mu}^B)_{\ b}^a=A_{\ b,\mu}^a/\sqrt{2}$. (The
factor $1/\sqrt{2}$ comes from a normalization convention.)
\par  
The covariant derivative acting on the quark field is
\be \lb{e2}
D_{\ b,\mu}^a=\delta_{\ b}^a\partial_{\mu}+i\frac{g}{\sqrt{2}}
A_{\ b,\mu}^a,
\ee
where $g$ is the (dimensionless) coupling constant of the theory.
\par
As is the case of connections, the gluon field does not transform 
covariantly under gauge transformations; it obtains an inhomogeneous 
part that breaks covariance. It is possible, however, to construct
a new field that has covariance properties. This is the gluon
field strength $F$ (the analog of the curvature tensor of
differential geometry), given by
\be \lb{e3}
F_{\ b,\mu\nu}^a=\partial_{\mu}A_{\ b,\nu}^a-\partial_{\nu}A_{\ b,\mu}^a
+i\frac{g}{\sqrt{2}}(A_{\ c\mu}^aA_{\ b\nu}^c-A_{\ c\nu}^aA_{\ b\mu}^c).
\ee
Its transformation law is
\be \lb{e4}
F_{\ b,\mu\nu}^a(x)\longrightarrow F_{\ \ b,\mu\nu}^{'a}(x)=
\Omega_{\ c}^a(x)\ F_{\ d,\mu\nu}^c(x)\ \Omega_{\ \ b}^{\dagger d}(x).
\ee
\par
With the aid of the covariant fields, one can easily construct the
gauge-invariant Lagrangian density of the theory:
\be \lb{e5}
\mathcal{L}=-\frac{1}{4}F_{\ b,\mu\nu}^{a}F_{\ a}^{b,\mu\nu}
+\overline{\psi}_ai\gamma^{\mu}D_{\ b,\mu}^a\psi^b
-m\overline\psi_a\psi^a,
\ee
where $m$ is the mass parameter of the quark fields and $\gamma^{\mu}$
are the Dirac matrices (the spinor indices being omitted).
\par

\section{Parallel transport} \lb{s3}

Let us consider the ordinary two-point Green's function of a quark 
field:
\be \lb{e6}
G_{\ b,\alpha\beta}^a(x-y)=<0|T\Big(\psi_{\alpha}^a(x)
\overline\psi_{b,\beta}(y)\Big)|0>=\langle \psi_{\alpha}^a(x)
\overline\psi_{b,\beta}(y)\rangle,
\ee
where the first expression in the right-hand-side represents the
vacuum expectation value of the chronological product of the
two quark fields, while the second expression corresponds to 
the notation used in the path-integral formalism; we recall that, 
in the latter formalism, ordinary products of functions actually 
represent chronological products of the operator formalism.
\par
Considering the gauge transformation law of the quark fields as 
given in Eqs. (\rf{e1}), we immediately deduce that the above 
Green's function is not invariant under local gauge transformations, 
since the vacuum state and the Lagrangian density (\rf{e5}) are 
gauge invariant. The consequence of this property can only be the 
vanishing of the Green's function for general values of the 
coordinates $x$ and $y$ \cite{Collins:1984xc}. This would mean the 
impossibility of treating perturbatively noninvariant quantities. 
The resolution of this difficulty comes from adding into the Lagrangian 
density (\rf{e5}) gauge-fixing terms and ghost fields 
\cite{Faddeev:1967fc}. The classical local gauge invariance is then 
replaced by a quantized global invariance, the BRST symmetry 
\cite{Becchi:1975nq,Tyutin:1975qk}. However, in gauge-invariant 
Green's functions the gauge-fixing and  ghost-field 
contributions generally cancel each other and one may continue 
reasoning, at a primary level, with the classical gauge transformations.
\par
To construct gauge-invariant Green's functions, one should
relate in some way gauge transformations at the point $x$ with those
at the point $y$. This is generally done by devising an operation of 
parallel transport, for instance from point $x$ to point $y$. This 
operation is usually realized with the aid of the connection. 
By considering the parallel transport along an oriented curve $C_{yx}$, 
its representative function, which we denote $U(C_{yx};y,x)$, takes 
the form of a path-ordered phase factor of the gluon field:
\be \lb{e7}
U_{\ b}^a(C_{yx};y,x)=\Big(Pe^{\displaystyle{-ig\int_x^ydz^{\mu}
\ T^BA_{\ \mu}^B(z)}}\Big)_{\ b}^a\ ,
\ee
where $P$ represents the path-ordering operation, meaning that the
gluon fields are ordered according to their position on the curve
$C_{yx}$. The above exponential can also be decomposed into
products of other exponentials, each defined successively on a 
smaller part of the curve $C_{yx}$. A more explicit expression
of $U$ can be obtained by a series expansion of the exponential
in terms of the coupling constant,
in which path ordering is taken into account. Parametrizing the 
function $z$ on the curve with a 
parameter $\lambda$ varying between 0 and 1, such that $z(0)=x$ and 
$z(1)=y$, and defining 
$z'(\lambda)=\frac{\partial z(\lambda)}{\partial \lambda}$, one
obtains
\bea \lb{e8}    
U_{\ b}^a(C_{yx};y,x)&=&\delta_{\ b}^a+\Big(\frac{-ig}{\sqrt{2}}\Big)
\int_0^1d\lambda_1\ z^{\prime\mu_1}(\lambda_1)\ 
A_{\ b,\mu_1}^a(\lambda_1) \nonumber\\
& &+\sum_{n=2}^{\infty}\Big(\frac{-ig}{\sqrt{2}}\Big)^n
\int_0^1d\lambda_1\cdots d\lambda_n\ 
\theta(\lambda_1-\lambda_2)\cdots\theta(\lambda_{n-1}-\lambda_n) 
\nonumber\\
& &\times z^{\prime\mu_1}(\lambda_1)\cdots z^{\prime\mu_n}(\lambda_n)
\ A_{\ c_1,\mu_1}^a(\lambda_1)A_{\ c_2,\mu_2}^{c_1}(\lambda_2)
\cdots A_{\ b,\mu_n}^{c_{n-1}}(\lambda_n),
\eea
where the abbreviation $A(z(\lambda))=A(\lambda)$ has been used.
\par
Under a gauge transformation, $U$ behaves as a covariant bilocal
field, the transformation depending only on its end points
$x$ and $y$, independently of the form of the curve $C_{yx}$
\cite{Corrigan:1978zg}:
\be \lb{e9}
U_{\ b}^a(C_{yx};y,x)\ \longrightarrow\  
\Omega_{\ c}^a(y)\ U_{\ d}^c(C_{yx};y,x)\ \Omega_{\ \ b}^{\dagger d}(x).
\ee
\par
By applying the operator $U$ on a quark field at point $x$, its variation
at $x$ cancels the one coming from the quark field and the whole
object transforms as belonging to the fundamental representation at
point $y$:
\be \lb{e10}
U_{\ b}^a(C_{yx};y,x)\ \psi^b(x)\ \longrightarrow\  
\Omega_{\ c}^a(y)\ U_{\ b}^c(C_{yx};y,x)\ \psi^b(x). 
\ee
Finally, by applying the previous object from the left on an antiquark
field, the $y$-dependent variation of the latter is canceled and the 
whole quantity becomes invariant under gauge transformations:
\be \lb{e11}
\overline \psi_{a}(y)\ U_{\ b}^a(C_{yx};y,x)\ \psi^b(x)\ \longrightarrow\  
\overline \psi_{a}(y)\ U_{\ b}^a(C_{yx};y,x)\ \psi^b(x). 
\ee
Considering this object in the path-integral formalism, or taking
the chronological product of the fields once their color indices 
are fixed, allows one to define, up to a normalization factor, a 
gauge-invariant quark Green's function. The latter depends, in addition to 
the coordinates of the quark fields, on the line $C_{yx}$ that appears 
in the definition of the parallel transport operation from point $x$ to 
point $y$. We shall denote the resulting Green's function $S(x,y;C_{yx})$; 
it no longer depends on color indices and is given by
\be \lb{e12}
S_{\alpha\beta}(x,y;C_{yx})=-\frac{1}{N_c}\langle 
\overline \psi_{a,\beta}(y)\ U_{\ b}^a(C_{yx};y,x)\ \psi_{\alpha}^b(x) 
\rangle.
\ee  
\par 

\section{Wilson loops} \lb{s4}

Another gauge-invariant useful object is constructed with the 
sole parallel transport operator (\rf{e7}). Considering in the latter 
the case of a closed line $C_{xx}$, obtained from $C_{yx}$ by the
limit $y\rightarrow x$ along a nontrivial curve, and then taking
the trace of the color indices at the matching point, one
clearly obtains from Eq. (\rf{e9}) a gauge-invariant result:
\be \lb{e13}
U_{\ a}^a(C_{xx};x,x)\ \longrightarrow\  
U_{\ a}^a(C_{xx};x,x) .
\ee
The line $C_{xx}$ being closed and the trace having been taken,
the point $x$  no longer plays a particular role in that object
and hence can be removed from the representation of the operator,
where only the closed contour may appear. The resulting object is
called a Wilson loop:
\be \lb{e14}
\Phi(C)=\frac{1}{N_c}\ \mathrm{tr}\ \Big(Pe^{\displaystyle{-ig\oint_C 
dz^{\mu}\ T^BA_{\ \mu}^B(z)}}\Big)\ =\ \frac{1}{N_c}\ U(C)_{\ a}^a.
\ee 
Its vacuum expectation value \cite{Wilson:1974sk} will be denoted by
\be \lb{e15}
W(C)=\langle \Phi(C)\rangle .
\ee
\par
The Wilson loop provides a simple criterion for the confinement of
quarks. In the case of a static pair of a quark and an antiquark, 
considered as the extreme limit of heavy quarks, which remain fixed 
in position space and which are observed at equal times, the whole 
dynamics of the system in the color singlet state can be described 
by means of a Wilson loop average (\rf{e15}) along a
rectangular contour in a plane, where one of the sides represents
the fixed distance $R$ between the quark and the antiquark and the
other side the interval $T$ of the evolved time (see Fig. \rf{f1}).
\bfg
\vspace*{0.5 cm}
\bc
\begin{picture}(0,0)%
\includegraphics{f1.pstex}%
\end{picture}%
\setlength{\unitlength}{2960sp}%
\begingroup\makeatletter\ifx\SetFigFont\undefined%
\gdef\SetFigFont#1#2#3#4#5{%
  \reset@font\fontsize{#1}{#2pt}%
  \fontfamily{#3}\fontseries{#4}\fontshape{#5}%
  \selectfont}%
\fi\endgroup%
\begin{picture}(3927,2466)(2986,-4030)
\put(6826,-1786){\makebox(0,0)[lb]{\smash{{\SetFigFont{9}{10.8}{\familydefault}{\mddefault}{\updefault}{\color[rgb]{0,0,0}$x_1'$}%
}}}}
\put(5026,-1711){\makebox(0,0)[lb]{\smash{{\SetFigFont{9}{10.8}{\familydefault}{\mddefault}{\updefault}{\color[rgb]{0,0,0}$T$}%
}}}}
\put(6826,-3961){\makebox(0,0)[lb]{\smash{{\SetFigFont{9}{10.8}{\familydefault}{\mddefault}{\updefault}{\color[rgb]{0,0,0}$x_2'$}%
}}}}
\put(3151,-3961){\makebox(0,0)[lb]{\smash{{\SetFigFont{9}{10.8}{\familydefault}{\mddefault}{\updefault}{\color[rgb]{0,0,0}$x_2$}%
}}}}
\put(3001,-2836){\makebox(0,0)[lb]{\smash{{\SetFigFont{9}{10.8}{\familydefault}{\mddefault}{\updefault}{\color[rgb]{0,0,0}$R$}%
}}}}
\put(3151,-1786){\makebox(0,0)[lb]{\smash{{\SetFigFont{9}{10.8}{\familydefault}{\mddefault}{\updefault}{\color[rgb]{0,0,0}$x_1$}%
}}}}
\end{picture}%

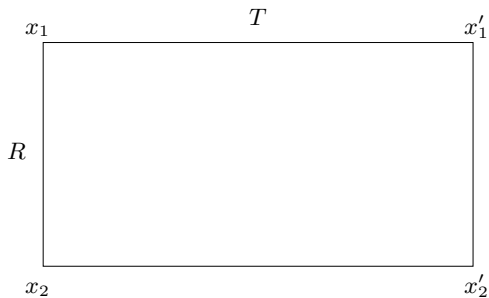
\caption{Rectangular contour of a Wilson loop representing the 
evolution of a static quark-antiquark system.}
\lb{f1}
\ec
\efg
For large separation distances and large time intervals, the 
following relationship can be shown \cite{Brown:1979ya}: 
\be \lb{e16}
W(C_{\mathrm{rect.}})\ \sim\ e^{\displaystyle{-iV(R)T}},
\ee
where $V(R)$ is the static potential energy corresponding to the 
force that is exerted between the quark and the antiquark. However, 
$W(C_{\mathrm{rect.}})$ can be analytically calculated 
in lattice QCD \cite{Wilson:1974sk,Kogut:1982ds,Kaku:1993ym} in 
the strong-coupling approximation, yielding
\be \lb{e17}  
W(C_{\mathrm{rect.}})\ \sim\ e^{\displaystyle{-i\sigma RT}},  
\ee
where $\sigma$ is a positive constant with dimension energy/length. 
Comparing the two expressions (\rf{e16}) and (\rf{e17}), one deduces
\be \lb{e18}
V(R)=\sigma R,
\ee
which shows that the interquark static potential energy is a 
linearly increasing function of the distance. This is sufficient to 
bind the quark and the antiquark together and to prevent their 
separation to infinite distances. Since the product $RT$ represents
the area of the rectangular contour, the above result is called the 
area law of confinement. Assuming that the previous result remains
in general true in the continuum limit of lattice QCD, one ends
up with the conclusion that QCD in the continuum should confine
quarks, in accordance with the observational facts. The constant
$\sigma$ is called the string tension; the numerical value of
$\sqrt{\sigma}$ lies within the bounds 420--480 MeV (with units
where $\hbar=c=1$) \cite{Bali:2000gf}. 
\par
The large-time behavior in the static limit of the Wilson loop average 
determines therefore the interquark potential energy. Because the latter 
is a quantity defined mainly in nonrelativistic theories, it appears
that the Wilson loop itself should be the quantity that would play
the role of potentials in the general case. Being gauge invariant,
the information coming from it about the dynamics of quarks would
have an unambiguous interpretation. In the case of general simple
contours, the area law obtained for the rectangular contour would 
be replaced by the area of the surface, supported by the contour, 
having a minimal value (called a minimal surface)
\cite{Wilson:1974sk}. Wilson loops are often used in relativistic
approaches to bound state problems \cite{Eichten:1980mw,
Brambilla:1993zw,Brambilla:2000gk,Dubin:1994vn,Simonov:1987rn}. 
\par
Wilson loop averages can be considered as functionals of the contour
$C$ from which they are defined. They may therefore receive a 
geometric interpretation or be subjected to a geometric analysis.
Properties of Wilson loops have been thoroughly studied, starting with
works by Polyakov \cite{Polyakov:1980ca} and Makeenko and Migdal 
\cite{Makeenko:1979pb,Makeenko:1980wr,Makeenko:1980vm,Migdal:1984gj}. 
\par
The analysis is based upon functional derivations acting on the
path-ordered phase factor (the parallel transport operator), which 
may be submitted to local deformations of its line. 
Two main equations are obtained, the Bianchi identity and the loop 
equation, the latter resulting from the equation of motion operator 
of the gluon field. Owing to the complexity of the loop equation,  
obtaining an exact nonperturbative solution of 
it displaying confinement has not been possible. However, Makeenko 
and Migdal showed that,
for large simple contours, the loop equations have asymptotic 
solutions that satisfy the area law of confinement 
\cite{Makeenko:1980wr}, obtained independently in lattice QCD 
\cite{Wilson:1974sk,Kogut:1982ds,Kaku:1993ym}. In perturbation theory, 
renormalization of Wilson loops was shown in 
\cite{Dotsenko:1979wb,Brandt:1981kf}.
\par    
The loop equation was also analyzed in two spacetime dimensions
by Kazakov and Kostov \cite{Kazakov:1980zi,Kazakov:1980zj}, who
could exactly solve the equation for many kinds of contour.
In particular, for simple contours, the solution is given by the 
exponential of the area of the surface enclosed by them: The area law 
is thus explicitly satisfied. The same results were also obtained by
Brali\'c \cite{Bralic:1980ra}, who used the non-Abelian version of 
the Stokes theorem for the calculation of the Wilson loops. Actually, 
the main simplification in two dimensions arises from the fact that 
the QCD coupling constant $g$ has now a dimension of mass ($\hbar=c=1$) 
and its square has the dimension of the string tension that appears in
the area law; this feature naturally drags the solutions to the 
area law, which is also obtained in perturbation theory in the axial
gauge. 
\par
A general survey of questions related to loop equations and gauge
theories can be found in \cite{Makeenko:1999hq}.
Properties of Wilson loops saturated with minimal surfaces are 
studied in \cite{Jugeau:2003df}. It has been shown that the minimal
surface is the only type of surface that satisfies the Bianchi
identity. This implies that fluctuations of surfaces around a minimal 
surface violate in general the Bianchi identity 
\cite{Makeenko:1980vm,Jugeau:2003df}. 
The violation of the Bianchi identity is a signal of the presence
of a color monopole current. However, the Bianchi identity
plays an important role in the verification of some consistency 
conditions of the loop formalism, in particular for ensuring the 
commutativity of two successive functional derivatives 
\cite{Jugeau:2003df}. It has also been shown that, by neglecting the
short-distance (i.e., perturbative) interactions and renormalizing
the coupling constant $g^2N_c$ quadratically with respect to a
short-distance regulator, one ends up, for simple contours, with a 
solution to the loop equation that is identical to a
minimal surface \cite{Jugeau:2003df}. This provides another 
verification of the asymptotic solution obtained in 
\cite{Makeenko:1980wr}. (In perturbation theory, $g^2N_c$ vanishes
with the short-distance regulator as the inverse of a logarithm.)    
\par

\section{Polygonal lines for parallel transport} \lb{s5}

Although loop equations have not been solved exactly or analytically
in four spacetime dimensions, the past investigations provide us with 
a general idea about the main features of Wilson loop vacuum averages.
For simple large contours, minimal surfaces are asymptotic solutions,
in accordance with the area law obtained in lattice QCD in the 
strong-coupling regime. For small contours, ordinary perturbation theory 
should be applicable for their evaluation. Adopting this qualitative 
basis, one can go further, using Wilson loops in dynamical equations 
where they play the role of gauge-invariant potentials or kernels. 
Since most experimental observations related with QCD concern hadrons,
which involve quarks as main degrees of freedom, the study of the
properties of gauge-invariant quark Green's functions appears as a
natural step for such an investigation. We already defined in Eq.
(\rf{e12}) the two-point gauge-invariant quark Green's function
(2PGIQGF), which is a functional of the line defining the parallel
transport from the quark field to the antiquark field. 
\par
At this point the question arises as to the impact of the type of
phase factor line on physical quantities: Do the latter 
depend on the choice made of the line followed by the parallel transport
operation? A general answer to this question has been given in Ref.
\cite{Jugeau:2003df} (end of Appendix A), in the simplified case 
of a static quark-antiquark system evolving in time, within the
framework of the saturation of Wilson loop averages by minimal
surfaces. One compares the evolution law of a system defined with
a straight line segment for the phase factor line with that of 
a system defined with an arbitrary (continuous) line. The physical 
observable is represented here by the energy content of the system.
The latter is explicitly exhibited through an exponential factor
when the evolution time interval $T$ tends to infinity (see Eq.
(\rf{e16})). The equations of the minimal surfaces allow us to 
study in detail this limit: Both minimal surfaces yield the same 
energy dependence, governed actually by that of the straight line 
segment and the rectangular Wilson loop (Fig. \rf{f1}). The line
dependencies are factored out within the wave functional of each
system. The above analysis displays the effect the choice of the
phase factor line may have on a physical system under study: It
affects the wave functional of the system, but it has no influence
on physical quantities. The case of nonstatic quarks does not seem 
to modify in general the previous conclusion and we adopt the 
latter in the forthcoming part of our investigation. 
\par
Since the simplest choice for a phase factor line is the straight
line segment, one would be tempted to stick to Green's functions
defined with such lines. This is, however, not possible as a single
stage, because the equations of motion couple, through interaction 
terms, Green's functions with different lines. A single choice of a 
rigid line does not lead to a closed system of equations. The 
situation here is very similar to what happens with the Dyson--Schwinger 
equations, which consist of an infinite number of coupled equations. 
In the present case, in view of the complexity of all types of line, 
it is natural to seek classes of line that might form a closed system 
under the equations of motion. This problem  actually has a positive 
answer: Polygonal lines, made of a succession of straight line segments, 
joined to each other, do form a closed set, at least for studies 
related to quark Green's functions \cite{Sazdjian:2007ng}.  
\par
Polygonal lines have several advantages worth mentioning: 
(a) They can be classified according to the number of 
segments they contain; accordingly, the same classification appears 
also at the level of the corresponding Green's functions. (b) Their
basic ingredients, which are the straight line segments, are Lorentz
invariant in form, which also reflects itself on the previous 
classification scheme. (c) The straight line segment has a well-defined, 
unambiguous limit when its two end points approach each other: It shrinks 
to a point. (d) Polygonal lines form a complete set for the present 
problem, in the sense that all equations that are derived will be closed
within the category of polygonal line; no other types of line will 
be needed to complete the description.      
We shall henceforth consider for the parallel transport operator the 
class of line having a polygonal structure in spacetime.
\par
One of the key ingredients in our study is a functional variation
formula, first established by Mandelstam \cite{Mandelstam:1968hz},
which results when local deformations of the line $C_{yx}$ are 
considered. If the line $C_{yx}$ is deformed at all its points by
infinitesimal variations $\delta z(\lambda)$, including displacements 
of its end points, then the parallel transport operator undergoes the 
variation
\cite{Mandelstam:1968hz,Nambu:1978bd,Corrigan:1978zg,Polyakov:1980ca,
Durand:1979sw}
\bea \lb{e23}
\delta U_{\ b}^a(1,0)&=&-i\frac{g}{\sqrt{2}}\delta z^{\alpha}(1)
A_{\ c,\alpha}^a(1)U_{\ b}^c(1,0)
+i\frac{g}{\sqrt{2}}U_{\ c}^a(1,0)A_{\ b,\alpha}^c(0)
\delta z^{\alpha}(0)\nonumber \\
& &+i\frac{g}{\sqrt{2}}\int_0^1d\lambda\ U_{\ c}^a(1,\lambda)
z^{\prime\beta}(\lambda)F_{\ d,\beta\alpha}^c(\lambda)
\delta z^{\alpha}(\lambda)U_{\ b}^d(\lambda,0),
\eea
where the abbreviation $U(\lambda_2,\lambda_1)$ has been used to 
designate the operator $U$ corresponding to the line going from 
$z(\lambda_1)$ to $z(\lambda_2)$ along the curve $C_{yx}$, with 
$z(0)=x$ and $z(1)=y$. 
This formula exhibits the relationship between dynamical operations,
such as the insertion of the gluon field strength inside the 
phase factor, and geometric operations, such as line deformations.
\par
We now consider for the line $C_{yx}$ an oriented straight line segment 
going from $x$ to $y$ and designate by $U(y,x)$ the corresponding
parallel transport operator. A displacement of one end point of the 
rigid segment in form, while the other end point remains fixed, generates 
 a displacement of the interior points of the segment. This defines a 
rigid path displacement. By parametrizing the interior points of the segment 
with a linear parameter $\lambda$ varying between $0$ and $1$, such that 
$z(\lambda)=\lambda y+(1-\lambda)x$,
the rigid path derivative operations with respect to $y$ or $x$ yield
\be \lb{e24}
\frac{\partial U_{\ b}^a(y,x)}{\partial y^{\alpha}}=
-i\frac{g}{\sqrt{2}}A_{\ c,\alpha}^a(y)U_{\ b}^c(y,x)
+\frac{\bar\delta U_{\ b}^a(y,x)}{\bar\delta y^{\alpha +}},
\ee
\be \lb{e25}
\frac{\partial U_{\ b}^a(y,x)}{\partial x^{\alpha}}=
+i\frac{g}{\sqrt{2}}U_{\ c}^a(y,x)A_{\ b,\alpha}^c(x)
+\frac{\bar\delta U_{\ b}^a(y,x)}{\bar\delta x^{\alpha -}},
\ee
where the second terms on the right-hand sides represent the
contributions of the interior points of the segment and are 
given by the following expressions:
\bea 
\lb{e26}  
& &\frac{\bar\delta U_{\ b}^a(y,x)}{\bar\delta y^{\alpha +}}=
i\frac{g}{\sqrt{2}}(y-x)^{\beta}\int_0^1d\lambda\lambda 
U_{\ c}^a(y,z(\lambda))F_{\ d,\beta\alpha}^c(z(\lambda))
U_{\ b}^d(z(\lambda),x),\\
\lb{e27}
& &\frac{\bar\delta U_{\ b}^a(y,x)}{\bar\delta x^{\alpha -}}=
i\frac{g}{\sqrt{2}}(y-x)^{\beta}\int_0^1d\lambda(1-\lambda) 
U_{\ c}^a(y,z(\lambda))F_{\ d,\beta\alpha}^c(z(\lambda))
U_{\ b}^d(z(\lambda),x).
\eea
The superscript $+$ or $-$ on the derivative variable takes account 
of the orientation on the segment and specifies, in the case of
joined segments, the segment on which the derivative acts.
\par
In the case of a polygonal contour $C_n$, with $n$ segments and 
$n$ junction points $x_1$, $x_2$, $\ldots$, $x_n$, the vacuum average
of the Wilson loop will be designated by $W_n=W(x_n,x_{n-1},\ldots,x_1)$
(see Fig. \rf{f2}).
\bfg
\vspace*{0.5 cm}
\bc
\begin{picture}(0,0)%
\includegraphics{f2.pstex}%
\end{picture}%
\setlength{\unitlength}{2486sp}%
\begingroup\makeatletter\ifx\SetFigFont\undefined%
\gdef\SetFigFont#1#2#3#4#5{%
  \reset@font\fontsize{#1}{#2pt}%
  \fontfamily{#3}\fontseries{#4}\fontshape{#5}%
  \selectfont}%
\fi\endgroup%
\begin{picture}(3945,3276)(3766,-6439)
\put(5266,-6361){\makebox(0,0)[lb]{\smash{{\SetFigFont{9}{10.8}{\familydefault}{\mddefault}{\updefault}{\color[rgb]{0,0,0}$x_1$}%
}}}}
\put(7696,-5956){\makebox(0,0)[lb]{\smash{{\SetFigFont{9}{10.8}{\familydefault}{\mddefault}{\updefault}{\color[rgb]{0,0,0}$x_5$}%
}}}}
\put(6841,-3481){\makebox(0,0)[lb]{\smash{{\SetFigFont{9}{10.8}{\familydefault}{\mddefault}{\updefault}{\color[rgb]{0,0,0}$x_4$}%
}}}}
\put(5221,-3346){\makebox(0,0)[lb]{\smash{{\SetFigFont{9}{10.8}{\familydefault}{\mddefault}{\updefault}{\color[rgb]{0,0,0}$x_3$}%
}}}}
\put(5626,-4786){\makebox(0,0)[lb]{\smash{{\SetFigFont{9}{10.8}{\familydefault}{\mddefault}{\updefault}{\color[rgb]{0,0,0}$W_5$}%
}}}}
\put(3781,-4831){\makebox(0,0)[lb]{\smash{{\SetFigFont{9}{10.8}{\familydefault}{\mddefault}{\updefault}{\color[rgb]{0,0,0}$x_2$}%
}}}}
\end{picture}%

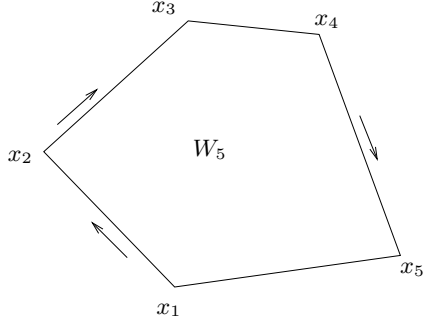
\caption{A Wilson loop along a polygonal contour with five sides.}
\lb{f2}
\ec
\efg
\par

\section{Quark Green's functions} \lb{s6}

By specializing to a polygonal line, the 2PGIQGF [Eq. (\rf{e12})] 
can be classified according 
to the number of segments contained in the line.
For a polygonal line composed of $n$ segments and $(n-1)$ junction 
points, the corresponding Green's function will be designated by 

\be \lb{e29}
S_{(n)}(x,x';t_{n-1},\ldots,t_1)=-\frac{1}{N_c}\langle 
\overline\psi(x')U(x',t_{n-1})U(t_{n-1},t_{n-2})\ldots U(t_1,x)
\psi(x)\rangle,
\ee
where each phase factor $U$ is along a straight line segment 
indicated by its two end-point coordinates. (Spinor indices are omitted 
and the color indices are implicitly summed.) The simplest 
such function is $S_{(1)}$, having a phase factor along a single 
straight line segment:
\be \lb{e30} 
S_{(1)}(x,x')\equiv S(x,x')=-\frac{1}{N_c}\langle 
\overline\psi(x')U(x',x)\psi(x)\rangle.
\ee
(We shall generally omit the index 1 from that function.) Figure 
\rf{f3} represents pictorially two different Green's functions.
\bfg
\vspace*{0.5 cm}
\bc
\begin{picture}(0,0)%
\includegraphics{f3.pstex}%
\end{picture}%
\setlength{\unitlength}{2368sp}%
\begingroup\makeatletter\ifx\SetFigFont\undefined%
\gdef\SetFigFont#1#2#3#4#5{%
  \reset@font\fontsize{#1}{#2pt}%
  \fontfamily{#3}\fontseries{#4}\fontshape{#5}%
  \selectfont}%
\fi\endgroup%
\begin{picture}(7162,2590)(2161,-5243)
\put(2176,-4261){\makebox(0,0)[lb]{\smash{{\SetFigFont{9}{10.8}{\familydefault}{\mddefault}{\updefault}{\color[rgb]{0,0,0}$x$}%
}}}}
\put(4351,-4261){\makebox(0,0)[lb]{\smash{{\SetFigFont{9}{10.8}{\familydefault}{\mddefault}{\updefault}{\color[rgb]{0,0,0}$x'$}%
}}}}
\put(6976,-4561){\makebox(0,0)[lb]{\smash{{\SetFigFont{9}{10.8}{\familydefault}{\mddefault}{\updefault}{\color[rgb]{0,0,0}$x$}%
}}}}
\put(9151,-4561){\makebox(0,0)[lb]{\smash{{\SetFigFont{9}{10.8}{\familydefault}{\mddefault}{\updefault}{\color[rgb]{0,0,0}$x'$}%
}}}}
\put(7426,-3361){\makebox(0,0)[lb]{\smash{{\SetFigFont{9}{10.8}{\familydefault}{\mddefault}{\updefault}{\color[rgb]{0,0,0}$t_1$}%
}}}}
\put(8701,-2836){\makebox(0,0)[lb]{\smash{{\SetFigFont{9}{10.8}{\familydefault}{\mddefault}{\updefault}{\color[rgb]{0,0,0}$t_2$}%
}}}}
\put(3001,-5161){\makebox(0,0)[lb]{\smash{{\SetFigFont{9}{10.8}{\familydefault}{\mddefault}{\updefault}{\color[rgb]{0,0,0}$S_{(1)}$}%
}}}}
\put(7951,-5161){\makebox(0,0)[lb]{\smash{{\SetFigFont{9}{10.8}{\familydefault}{\mddefault}{\updefault}{\color[rgb]{0,0,0}$S_{(3)}$}%
}}}}
\end{picture}%

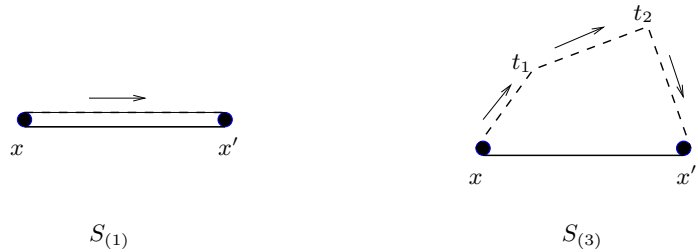
\caption{Pictorial representation of the Green's functions $S_{(1)}$
and $S_{(3)}$. Dashed lines represent the phase factor lines. Full
circles represent the position of the quark and antiquark fields, and
full lines represent the contraction operation between them.}  
\lb{f3}
\ec
\efg
\par
For the internal parts of rigid path derivatives, we have 
definitions of the type
\be \lb{e31}
\frac{\bar\delta S_{(n)}(x,x';t_{n-1},\ldots,t_1)}{\bar\delta x^{\mu -}}
=-\frac{1}{N_c}\langle 
\overline\psi(x')U(x',t_{n-1})U(t_{n-1},t_{n-2})\cdots 
\frac{\bar\delta U(t_1,x)}{\bar\delta x^{\mu -}}
\psi(x)\rangle.
\ee
\par
The Green's functions satisfy the following equations
of motion concerning the quark field variables:
\bea \lb{e32}
& &(i\gamma.\partial_{(x)}-m)S_{(n)}(x,x';t_{n-1},\ldots,t_1)=
i\delta^4(x-x')W_n(x,t_{n-1},\ldots,t_1) \nonumber \\
& &\ \ \ \ \ +i\gamma^{\mu}\frac{\bar\delta S_{(n)}(x,x';t_{n-1},
\ldots,t_1)}{\bar\delta x^{\mu -}}.
\eea
\par
For $n=1$, we have
\be \lb{e33}
(i\gamma.\partial_{(x)}-m)\ S(x,x')=i\delta^4(x-x')+
i\gamma^{\mu}\frac{\bar\delta S(x,x')}{\bar\delta x^{\mu -}}.
\ee
Similar equations also hold with the variable $x'$.
A graphical representation of the equations of motion of
$S_{(1)}$ and $S_{(3)}$ is displayed in Fig. \rf{f4}. 
\par
\bfg
\vspace*{0.5 cm}
\bc
\begin{picture}(0,0)%
\includegraphics{f4.pstex}%
\end{picture}%
\setlength{\unitlength}{2763sp}%
\begingroup\makeatletter\ifx\SetFigFont\undefined%
\gdef\SetFigFont#1#2#3#4#5{%
  \reset@font\fontsize{#1}{#2pt}%
  \fontfamily{#3}\fontseries{#4}\fontshape{#5}%
  \selectfont}%
\fi\endgroup%
\begin{picture}(9402,3649)(886,-4718)
\put(3151,-1561){\makebox(0,0)[lb]{\smash{{\SetFigFont{11}{13.2}{\familydefault}{\mddefault}{\updefault}{\color[rgb]{0,0,0}$x$}%
}}}}
\put(4801,-1561){\makebox(0,0)[lb]{\smash{{\SetFigFont{11}{13.2}{\familydefault}{\mddefault}{\updefault}{\color[rgb]{0,0,0}$x'$}%
}}}}
\put(8401,-1561){\makebox(0,0)[lb]{\smash{{\SetFigFont{11}{13.2}{\familydefault}{\mddefault}{\updefault}{\color[rgb]{0,0,0}$x$}%
}}}}
\put(9976,-1561){\makebox(0,0)[lb]{\smash{{\SetFigFont{11}{13.2}{\familydefault}{\mddefault}{\updefault}{\color[rgb]{0,0,0}$x'$}%
}}}}
\put(3901,-1786){\makebox(0,0)[lb]{\smash{{\SetFigFont{11}{13.2}{\familydefault}{\mddefault}{\updefault}{\color[rgb]{0,0,0}$S$}%
}}}}
\put(9151,-1786){\makebox(0,0)[lb]{\smash{{\SetFigFont{11}{13.2}{\familydefault}{\mddefault}{\updefault}{\color[rgb]{0,0,0}$S$}%
}}}}
\put(3151,-4336){\makebox(0,0)[lb]{\smash{{\SetFigFont{11}{13.2}{\familydefault}{\mddefault}{\updefault}{\color[rgb]{0,0,0}$x$}%
}}}}
\put(4651,-4336){\makebox(0,0)[lb]{\smash{{\SetFigFont{11}{13.2}{\familydefault}{\mddefault}{\updefault}{\color[rgb]{0,0,0}$x'$}%
}}}}
\put(8626,-4336){\makebox(0,0)[lb]{\smash{{\SetFigFont{11}{13.2}{\familydefault}{\mddefault}{\updefault}{\color[rgb]{0,0,0}$x$}%
}}}}
\put(10051,-4336){\makebox(0,0)[lb]{\smash{{\SetFigFont{11}{13.2}{\familydefault}{\mddefault}{\updefault}{\color[rgb]{0,0,0}$x'$}%
}}}}
\put(3826,-4636){\makebox(0,0)[lb]{\smash{{\SetFigFont{11}{13.2}{\familydefault}{\mddefault}{\updefault}{\color[rgb]{0,0,0}$S_{(3)}$}%
}}}}
\put(5476,-1411){\makebox(0,0)[lb]{\smash{{\SetFigFont{11}{13.2}{\familydefault}{\mddefault}{\updefault}{\color[rgb]{0,0,0}$=$}%
}}}}
\put(7501,-1411){\makebox(0,0)[lb]{\smash{{\SetFigFont{11}{13.2}{\familydefault}{\mddefault}{\updefault}{\color[rgb]{0,0,0}$+$}%
}}}}
\put(5476,-4111){\makebox(0,0)[lb]{\smash{{\SetFigFont{11}{13.2}{\familydefault}{\mddefault}{\updefault}{\color[rgb]{0,0,0}$=$}%
}}}}
\put(7576,-4111){\makebox(0,0)[lb]{\smash{{\SetFigFont{11}{13.2}{\familydefault}{\mddefault}{\updefault}{\color[rgb]{0,0,0}$+$}%
}}}}
\put(6376,-3586){\makebox(0,0)[lb]{\smash{{\SetFigFont{11}{13.2}{\familydefault}{\mddefault}{\updefault}{\color[rgb]{0,0,0}$W_3$}%
}}}}
\put(8626,-1261){\makebox(0,0)[lb]{\smash{{\SetFigFont{11}{13.2}{\familydefault}{\mddefault}{\updefault}{\color[rgb]{0,0,0}$\times$}%
}}}}
\put(901,-4111){\makebox(0,0)[lb]{\smash{{\SetFigFont{11}{13.2}{\familydefault}{\mddefault}{\updefault}{\color[rgb]{0,0,0}$(i\gamma . \partial_x-m)$}%
}}}}
\put(901,-1336){\makebox(0,0)[lb]{\smash{{\SetFigFont{11}{13.2}{\familydefault}{\mddefault}{\updefault}{\color[rgb]{0,0,0}$(i\gamma . \partial_x-m)$}%
}}}}
\put(8701,-3886){\makebox(0,0)[lb]{\smash{{\SetFigFont{11}{13.2}{\familydefault}{\mddefault}{\updefault}{\color[rgb]{0,0,0}$\times$}%
}}}}
\put(5851,-1786){\makebox(0,0)[lb]{\smash{{\SetFigFont{11}{13.2}{\familydefault}{\mddefault}{\updefault}{\color[rgb]{0,0,0}$i\delta^4(x-x')$}%
}}}}
\put(5851,-4561){\makebox(0,0)[lb]{\smash{{\SetFigFont{11}{13.2}{\familydefault}{\mddefault}{\updefault}{\color[rgb]{0,0,0}$i\delta^4(x-x')$}%
}}}}
\put(9301,-4636){\makebox(0,0)[lb]{\smash{{\SetFigFont{11}{13.2}{\familydefault}{\mddefault}{\updefault}{\color[rgb]{0,0,0}$S_{(3)}$}%
}}}}
\put(10051,-3136){\makebox(0,0)[lb]{\smash{{\SetFigFont{11}{13.2}{\familydefault}{\mddefault}{\updefault}{\color[rgb]{0,0,0}$t_2$}%
}}}}
\put(4651,-3061){\makebox(0,0)[lb]{\smash{{\SetFigFont{11}{13.2}{\familydefault}{\mddefault}{\updefault}{\color[rgb]{0,0,0}$t_2$}%
}}}}
\put(5701,-3136){\makebox(0,0)[lb]{\smash{{\SetFigFont{11}{13.2}{\familydefault}{\mddefault}{\updefault}{\color[rgb]{0,0,0}$t_1$}%
}}}}
\put(7126,-3136){\makebox(0,0)[lb]{\smash{{\SetFigFont{11}{13.2}{\familydefault}{\mddefault}{\updefault}{\color[rgb]{0,0,0}$t_2$}%
}}}}
\put(3451,-3061){\makebox(0,0)[lb]{\smash{{\SetFigFont{11}{13.2}{\familydefault}{\mddefault}{\updefault}{\color[rgb]{0,0,0}$t_1$}%
}}}}
\put(8776,-3136){\makebox(0,0)[lb]{\smash{{\SetFigFont{11}{13.2}{\familydefault}{\mddefault}{\updefault}{\color[rgb]{0,0,0}$t_1$}%
}}}}
\end{picture}%

\caption{Graphical representation of the equations of motion of
$S_{(1)}$ and $S_{(3)}$. The cross on the dashed lines represents  the
rigid path derivation; it is placed near the end point of the 
segment that is submitted to the derivation.}
\lb{f4}
\ec
\efg
\par
Multiplying Eq. (\rf{e32}) with $S(t_1,x)$ and integrating with 
respect to $x$, one obtains functional relations between various 
2PGIQGFs. A typical such relation is
\bea \lb{e34}
& &S_{(n)}(x,x';t_{n-1},\ldots,t_1)=S(x,x')\ 
W_{n+1}(x',t_{n-1},\ldots,t_1,x) \nonumber \\
& &\ \ \ \ \ +\Big(\frac{\bar\delta S(x,y_1)}
{\bar\delta y_1^{\alpha_1 +}}
+S(x,y_1)\frac{\bar\delta}{\bar\delta y_1^{\alpha_1 -}}\Big)\ 
S_{(n+1)}(y_1,x';t_{n-1},\ldots,t_1,x).
\eea
(Integrations on intermediate variables are implicit. Here, $y_1$ 
is an integration variable.) A graphical representation of this
equation for $n=3$ is given in Fig. \rf{f5}.
\par
\bfg
\vspace*{0.5 cm}
\bc
\begin{picture}(0,0)%
\includegraphics{f5.pstex}%
\end{picture}%
\setlength{\unitlength}{2763sp}%
\begingroup\makeatletter\ifx\SetFigFont\undefined%
\gdef\SetFigFont#1#2#3#4#5{%
  \reset@font\fontsize{#1}{#2pt}%
  \fontfamily{#3}\fontseries{#4}\fontshape{#5}%
  \selectfont}%
\fi\endgroup%
\begin{picture}(9886,2290)(736,-4643)
\put(751,-4336){\makebox(0,0)[lb]{\smash{{\SetFigFont{11}{13.2}{\familydefault}{\mddefault}{\updefault}{\color[rgb]{0,0,0}$x$}%
}}}}
\put(2326,-4336){\makebox(0,0)[lb]{\smash{{\SetFigFont{11}{13.2}{\familydefault}{\mddefault}{\updefault}{\color[rgb]{0,0,0}$x'$}%
}}}}
\put(3451,-4336){\makebox(0,0)[lb]{\smash{{\SetFigFont{11}{13.2}{\familydefault}{\mddefault}{\updefault}{\color[rgb]{0,0,0}$x$}%
}}}}
\put(1426,-4561){\makebox(0,0)[lb]{\smash{{\SetFigFont{11}{13.2}{\familydefault}{\mddefault}{\updefault}{\color[rgb]{0,0,0}$S_{(3)}$}%
}}}}
\put(2851,-4111){\makebox(0,0)[lb]{\smash{{\SetFigFont{11}{13.2}{\familydefault}{\mddefault}{\updefault}{\color[rgb]{0,0,0}$=$}%
}}}}
\put(4201,-3511){\makebox(0,0)[lb]{\smash{{\SetFigFont{11}{13.2}{\familydefault}{\mddefault}{\updefault}{\color[rgb]{0,0,0}$W_4$}%
}}}}
\put(6151,-3061){\makebox(0,0)[lb]{\smash{{\SetFigFont{11}{13.2}{\familydefault}{\mddefault}{\updefault}{\color[rgb]{0,0,0}$x$}%
}}}}
\put(8776,-3061){\makebox(0,0)[lb]{\smash{{\SetFigFont{11}{13.2}{\familydefault}{\mddefault}{\updefault}{\color[rgb]{0,0,0}$x$}%
}}}}
\put(5701,-4111){\makebox(0,0)[lb]{\smash{{\SetFigFont{11}{13.2}{\familydefault}{\mddefault}{\updefault}{\color[rgb]{0,0,0}$+$}%
}}}}
\put(8476,-4111){\makebox(0,0)[lb]{\smash{{\SetFigFont{11}{13.2}{\familydefault}{\mddefault}{\updefault}{\color[rgb]{0,0,0}$+$}%
}}}}
\put(6451,-3886){\makebox(0,0)[lb]{\smash{{\SetFigFont{11}{13.2}{\familydefault}{\mddefault}{\updefault}{\color[rgb]{0,0,0}$\times$}%
}}}}
\put(7051,-4561){\makebox(0,0)[lb]{\smash{{\SetFigFont{11}{13.2}{\familydefault}{\mddefault}{\updefault}{\color[rgb]{0,0,0}$S_{(4)}$}%
}}}}
\put(4201,-4561){\makebox(0,0)[lb]{\smash{{\SetFigFont{11}{13.2}{\familydefault}{\mddefault}{\updefault}{\color[rgb]{0,0,0}$S$}%
}}}}
\put(7951,-4261){\makebox(0,0)[lb]{\smash{{\SetFigFont{11}{13.2}{\familydefault}{\mddefault}{\updefault}{\color[rgb]{0,0,0}$x'$}%
}}}}
\put(10576,-4261){\makebox(0,0)[lb]{\smash{{\SetFigFont{11}{13.2}{\familydefault}{\mddefault}{\updefault}{\color[rgb]{0,0,0}$x'$}%
}}}}
\put(9226,-3811){\makebox(0,0)[lb]{\smash{{\SetFigFont{11}{13.2}{\familydefault}{\mddefault}{\updefault}{\color[rgb]{0,0,0}$\times$}%
}}}}
\put(9676,-4561){\makebox(0,0)[lb]{\smash{{\SetFigFont{11}{13.2}{\familydefault}{\mddefault}{\updefault}{\color[rgb]{0,0,0}$S_{(4)}$}%
}}}}
\put(5026,-4336){\makebox(0,0)[lb]{\smash{{\SetFigFont{11}{13.2}{\familydefault}{\mddefault}{\updefault}{\color[rgb]{0,0,0}$x'$}%
}}}}
\put(4951,-2986){\makebox(0,0)[lb]{\smash{{\SetFigFont{11}{13.2}{\familydefault}{\mddefault}{\updefault}{\color[rgb]{0,0,0}$t_2$}%
}}}}
\put(7876,-3136){\makebox(0,0)[lb]{\smash{{\SetFigFont{11}{13.2}{\familydefault}{\mddefault}{\updefault}{\color[rgb]{0,0,0}$t_2$}%
}}}}
\put(10501,-3136){\makebox(0,0)[lb]{\smash{{\SetFigFont{11}{13.2}{\familydefault}{\mddefault}{\updefault}{\color[rgb]{0,0,0}$t_2$}%
}}}}
\put(9751,-2536){\makebox(0,0)[lb]{\smash{{\SetFigFont{11}{13.2}{\familydefault}{\mddefault}{\updefault}{\color[rgb]{0,0,0}$t_1$}%
}}}}
\put(6601,-4261){\makebox(0,0)[lb]{\smash{{\SetFigFont{11}{13.2}{\familydefault}{\mddefault}{\updefault}{\color[rgb]{0,0,0}$y_1$}%
}}}}
\put(9226,-4261){\makebox(0,0)[lb]{\smash{{\SetFigFont{11}{13.2}{\familydefault}{\mddefault}{\updefault}{\color[rgb]{0,0,0}$y_1$}%
}}}}
\put(2176,-2986){\makebox(0,0)[lb]{\smash{{\SetFigFont{11}{13.2}{\familydefault}{\mddefault}{\updefault}{\color[rgb]{0,0,0}$t_2$}%
}}}}
\put(7126,-2536){\makebox(0,0)[lb]{\smash{{\SetFigFont{11}{13.2}{\familydefault}{\mddefault}{\updefault}{\color[rgb]{0,0,0}$t_1$}%
}}}}
\put(901,-2986){\makebox(0,0)[lb]{\smash{{\SetFigFont{11}{13.2}{\familydefault}{\mddefault}{\updefault}{\color[rgb]{0,0,0}$t_1$}%
}}}}
\put(3676,-2986){\makebox(0,0)[lb]{\smash{{\SetFigFont{11}{13.2}{\familydefault}{\mddefault}{\updefault}{\color[rgb]{0,0,0}$t_1$}%
}}}}
\end{picture}%

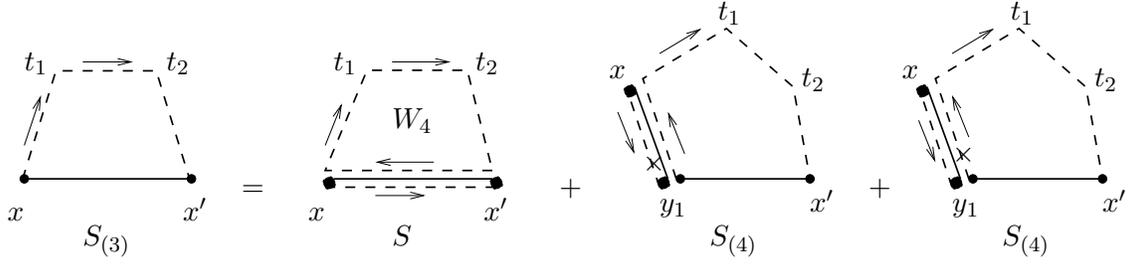
\caption{Graphical representation of the functional relation among
$S_{(3)}$, $S_{(1)}$, and $S_{(4)}$.}
\lb{f5}
\ec
\efg
\par
Since this equation is valid for any $n\ge 1$, one can use it again 
on its right-hand side for $S_{(n+1)}$. One therefore generates an 
iterative procedure that eliminates successively the higher index 
2PGIQGFs in favor of the lowest index one, $S_{(1)}$. Assuming that 
the terms rejected to infinity are negligible, one ends up with a 
series where only $S_{(1)}$ appears together with Wilson loop averages 
along polygonal contours with an increasing number of sides and rigid 
path derivatives along the segments. This result shows that, among the set 
of the 2PGIQGFs $S_{(n)}$, $n=1,2,\ldots$, it is only $S_{(1)}$, 
having a phase factor along one straight line segment, that is a 
genuine dynamical independent quantity. Higher index 2PGIQGFs could 
in principle be eliminated in terms of $S_{(1)}$ and polygonal Wilson 
loops and their rigid path derivatives.
\par
The above procedure is also repeated on the right-hand sides of the
equations of motion (\rf{e32}) and (\rf{e33}) to express the rigid 
path derivatives in terms of Wilson loop averages. The equation of
$S$ takes at the end the following form:
\bea \lb{e35}
& &(i\gamma.\partial_{(x)}-m)\,S(x,x')=i\delta^4(x-x')+
i\gamma^{\mu}\,\Big\{K_{1\mu -}(x',x)\,S(x,x')\nonumber \\
& &\ \ \ \ \ \ +K_{2\mu -}(x',x,y_1)\,S_{(2)}(y_1,x';x)\nonumber \\
& &\ \ \ \ \ \ +\sum_{n=3}^{\infty}K_{n\mu -}(x',x,y_1,\ldots,y_{n-1})\,
S_{(n)}(y_{n-1},x';x,y_1,\ldots,y_{n-2})\,\Big\},
\eea
where the kernels $K_n$ ($n=1,2,\ldots$) contain Wilson loop averages
along polygonal contours, which are at most $(n+1)$-sided, and the
2PGIQGF $S$ and its derivative. The total number of derivatives 
contained in $K_n$ is $n$.
Once the Wilson loop averages and the various derivatives have been
evaluated and the high-index $S_{(n)}$s have been expressed in terms
of $S$, Eq. (\rf{e35}) becomes an integro-\-differen\-tial equation 
in $S$, which is the primary unknown quantity to be solved. 
This equation is the analog of the self-energy Dyson--Schwinger
equation for ordinary Green's functions. Furthermore, the fact that
it has been obtained with the sole aid of polygonal lines for the 
parallel transport operator, without the need of other types of line, 
is an indication that polygonal lines form a complete set for the
study of the 2PGIQGFs.  
\par  
The various kernels that are present on the right-hand side of Eq.
(\rf{e35}) can be analyzed in terms of their behaviors at
large and short distances. It seems that the series is globally
perturbative with respect to the inverse of the number of sides 
of the polygonal contours, the first terms, having the least 
number of sides, being the dominant ones. The kernel $K_{1\mu -}$ 
is actually null for symmetry reasons; it is therefore the kernel 
$K_{2\mu -}$, corresponding to a triangular contour with two rigid 
path derivatives, that is expected to be the dominant term of the 
series.
\par
One of the mathematical difficulties of Eq. (\rf{e35}) comes from
the property that the kernels do not have a convolutive structure; 
this is related to the fact that Wilson loops act in each term on
all the points that are present in the corresponding expressions.
This is actually reminiscent of the difficulty of obtaining functional
inverses of gauge-invariant Green's functions because of their nonlocal
structure.
\par
 
\section{Two-dimensional QCD} \lb{s7}

The equations obtained in the previous sections also remain  valid 
in two spacetime dimensions and could be analyzed more easily in 
this case. Two-dimensional QCD in the large-$N_c$ limit 
\cite{'tHooft:1973jz,'tHooft:1974hx} provides a simplified framework 
for the study of the confinement properties, which are expected to 
be qualitatively similar to those of four dimensions. Wilson loop 
averages were explicitly calculated in two dimensions 
\cite{Kazakov:1980zi,Kazakov:1980zj,Bralic:1980ra}: For simple
contours they satisfy the area law. In that case, the second-order 
derivative of the logarithm of the Wilson loop average reduces to a 
two-dimensional delta function. Higher order derivatives give zero
in Eq. (\rf{e35}), since they act there on different segments of the 
polygonal contours. Overlapping self-intersecting surfaces, 
which give more complicated expressions, are assumed to give
negligible contributions, for the residual terms they produce are 
probably of zero weight under the integrations that are involved. 
\par
In the series of terms of Eq. (\rf{e35}) it is only the kernel
$K_2$ that survives and the integro-\-differen\-tial equation
takes the following expression \cite{Sazdjian:2010ku}:
\bea \lb{e36}
& &(i\gamma.\partial-m)S(x)=i\delta^2(x)
-\sigma\gamma^{\mu}(g_{\mu\alpha}g_{\nu\beta}-g_{\mu\beta}g_{\nu\alpha}) 
x^{\nu}x^{\beta}\nonumber \\
& &\ \ \ \ \ \times\left[\,\int_0^1d\lambda\,\lambda^2\,S((1-\lambda)x)
\gamma^{\alpha}S(\lambda x)
+\int_1^{\infty}d\xi\,S((1-\xi)x)\gamma^{\alpha}S(\xi x)\,\right], 
\eea
where $\sigma$ is the string tension.
\par
The above equation can be analyzed by first passing to momentum
space. Designating by $S(p)$ the Fourier transform of $S(x)$, one
can decompose it into Lorentz-invariant components:
\be \lb{e37}
S(p)=\gamma.pF_1(p^2)+F_0(p^2).
\ee
The solution of Eq. (\rf{e36}) can be searched for by using the 
analyticity properties of the 2PGIQGF.
It turns out that the equation can be solved exactly and in analytic 
form. The functions $F_1$ and $F_0$ are found to have an infinite 
number of branch cuts located on the positive real axis of $p^2$
(timelike region), starting at thresholds $M_1^2$, $M_2^2$, $\ldots$,
$M_n^2$, $\ldots$, with fractional power singularities equal to
$-3/2$. Their expressions, for complex $p^2$, are
\bea 
\lb{e38}
& &F_1(p^2)=-i\frac{\pi}{2\sigma}\sum_{n=1}^{\infty}\,b_n
\frac{1}{(M_n^2-p^2)^{3/2}},\\
\lb{e39}
& &F_0(p^2)=-i\frac{\pi}{2\sigma}\sum_{n=1}^{\infty}(-1)^{n+1}b_n
\frac{M_n}{(M_n^2-p^2)^{3/2}}.
\eea
The Green's function $S$ [Eq. (\rf{e37})] then takes the form
\be \lb{e40}
S(p)=-i\frac{\pi}{2\sigma}\sum_{n=1}^{\infty}\,b_n
\frac{(\gamma.p+(-1)^{n+1}M_n)}{(M_n^2-p^2)^{3/2}}.
\ee
The masses $M_n$ ($n=1,2,\ldots$) are positive, greater than the free 
quark mass $m,$ and ordered according to increasing values. For massless 
quarks they remain positive. The masses $M_n$ and the coefficients $b_n$,
the latter also being  positive, satisfy, for general $m$, an infinite 
set of algebraic equations that are solved numerically. Their asymptotic 
values, for large values of $n$ such that $n\gg m^2/(\pi\sigma)$, are       
\be \lb{e41}
M_n^2\simeq \pi n\sigma,\ \ \ \ \ \ \ \ 
b_n\simeq \frac{\sigma^2}{M_n+(-1)^nm}.
\ee
The functions $(M_n^2-p^2)^{-3/2}$ are defined with cuts starting from
their branch points and going to $+\infty$ on the real axis; they are
real below their branch points on the real axis down to $-\infty$.
\par
Expressions (\rf{e38}) and (\rf{e39}) represent weakly
converging series. The high-energy behavior of the functions $F_1$
and $F_0$ is obtained with a detailed study of the asymptotic tails
of the series and the use of the asymptotic behaviors of the parameters
$M_n$ and $b_n$ [Eqs. (\rf{e41})]. One finds that they behave 
asymptotically as in free field theories, which is here a trivial 
manifestation of asymptotic freedom \cite{Politzer:1976tv}, since in 
two dimensions and in the large-$N_c$ limit the coupling constant
does not undergo renormalization.
\par
In summary, the solution of Eq. (\rf{e36}) is nonperturbative
and infrared finite. The masses $M_n$ are dynamically generated,
since they do not exist in the Lagrangian of the theory. They could
be interpreted as dynamical masses of quarks with, however, the
following particular features. First, they are infinite in number. 
Second, they do not appear as poles in the Green's function but 
rather with stronger singularities. In $x$ space the latter do not 
produce finite plane waves at large distances and therefore quarks 
could not be observed as free asymptotic states. Nevertheless, the 
above singularities, being gauge invariant, should have physical 
significance and would show up in the infrared regions of physical 
processes involving quarks. 
\par
The fact that the singularities of the Green's function 
appear only in the timelike region of real $p^2$ is an indication
that the quark and gluon fields satisfy, even in the nonperturbative 
regime, the usual spectral properties of quantum field theory
\cite{Wightman:1956zz,Schweber:1961zz,'tHooft:1973pz}. Expression
(\rf{e40}) can be interpreted as fitting a generalized form
of the K\"all\'en--Lehmann representation 
\cite{Sazdjian:2007ng,Sazdjian:2010ku,Kallen:1952zz,Lehmann:1954xi}, 
where the denominator of the dispersive integral  now has a fractional 
power, while the spectral functions are saturated by an infinite 
series of dynamically generated single quark states with alternating 
parities. The latter still satisfy Lehmann's positivity conditions 
\cite{Lehmann:1954xi}.
\par    
Finally, the question may arise as to the dependence of the predictions
so far obtained on the choice of polygonal lines for the phase factor 
paths. According to our discussion at the beginning of Sec. \rf{s5}, 
physical quantities, such as masses and momentum space singularities,
should be insensitive to changes of line type and therefore should
be line independent.
\par 
 
\section{Conclusion} \lb{s8}

Gauge-invariant Green's functions are adequate tools for a 
systematic investigation of the nonperturbative properties of
QCD. Polygonal lines, supporting parallel transport in quark 
Green's functions, form a complete set of paths for the study
of the dynamical properties of the latter. An equation playing the 
same role as the self-energy Dyson--Schwinger equation for ordinary 
Green's functions has been obtained, in which the kernels are 
represented by Wilson loop averages along polygonal contours with 
rigid path derivatives on their segments. It is expected that the 
leading term in the series of kernels is provided by the Wilson loop 
along a triangular contour with two derivatives. 
\par
The application of this equation to two-dimensional QCD in the 
large-$N_c$ limit provides an exact non\-per\-tur\-bative analytic 
solution, not known from conventional approaches, that displays 
a dynamical generation of a series of massive quark states, with 
the characteristic feature that their singularities in 
momentum space are stronger than simple poles. 
\par
The consistency of the results obtained in two-dimensional QCD
is a positive test for the general approach presently developed
for investigations in four dimensions.
\par 

\vspace{0.5 cm}
\noindent
\textbf{Acknowledgements}
\par
This work was partially supported by EU I3HP Project ``Study of
Strongly Interacting Matter'' (acronym HadronPhysics3, Grant
Agreement No. 283286).
\par

\end{document}